\documentclass[aps,psfig,prd,preprintnumbers,amsmath,amssymb,superscriptaddress]{revtex4}
\usepackage{graphics}
\newcommand{\al}{\alpha}

\newcommand{\vect}{\mathbf}
\begin{document}
\title{Boundary Inflation in the Moduli Space Approximation}
\author{P.R.~Ashcroft}
\affiliation{Department of Applied Mathematics and Theoretical
Physics, Centre for Mathematical Sciences, University of Cambridge,
Wilberforce Road , Cambridge CB3 OWA, UK}
\author{C. van de Bruck}
\affiliation{Astrophysics Department, Oxford University, Keble
Road, Oxford OX1 3RH, UK}
\author{A.-C. Davis}
\affiliation{Department of Applied Mathematics and Theoretical
Physics, Centre for Mathematical Sciences, University of Cambridge,
Wilberforce Road , Cambridge CB3 OWA, UK}
\date{22 October 2003}
\begin{abstract}
The evolution of slow--roll inflation in a five--dimensional 
brane world model with two boundary branes and bulk scalar 
field is studied. Assuming that the inflationary scale is 
below the brane tension, we can employ the moduli space 
approximation to study the dynamics of the system. 
Detuning the brane tension results in a potential for the moduli
fields which we show will not support a period of slow--roll inflation. 
We then study an inflaton field, confined to the positive tension brane, 
to which the moduli fields are non--minimally coupled. We discuss in detail 
the two cases of $V(\chi) = \frac{1}{2} m^2 \chi^2$ and $V(\chi) = \lambda \chi^4$ and 
demonstrate that increasing the coupling results in spectra which are further 
away from scale--invariance and in an increase in the tensor mode production, 
while entropy perturbations are subdominant. Finally, we point out that the 
five--dimensional spacetime is unstable during inflation because the negative 
tension brane collapses.
\end{abstract}

\maketitle 

\vspace{0.5cm}
\noindent DAMTP-2003-111

\section{Introduction}
The discovery of branes in string theory has introduced a new 
class of model for the universe: the brane world. In this setup the universe is a 
three--dimensional object, embedded in a 
higher--dimensional spacetime. This has stimulated research 
along various avenues, in particular particle phenomenology and 
cosmology. (For reviews see [\ref{rubakov},\ref{branereview},\ref{langlois}]). 
In cosmology, there are different observational consequences which one can investigate. In particular, cosmological perturbations 
and varying constants have to be studied in detail. 
 
Current observations of the cosmic microwave background radiation (CMB) 
and large scale structures are remarkably consistent with the 
inflationary paradigm, according to which the universe underwent 
nearly exponential growth very early on. It is important 
to investigate how brane world scenarios modify the predictions 
of inflationary scenarios. A considerable amount of 
work has already been done in models of inflation with branes, in particular the high 
energy regime, where the expansion rate is directly proportional 
to the energy density, was subject to exhaustive investigations as well 
as models in which a bulk scalar field drives inflation (see, for example 
[\ref{wands1}]--[\ref{taylor}] and references therein). In this paper 
we will show that there are also new effects in the low--energy regime 
due to the presence of moduli fields which couple to matter on the branes. 
If the inflaton field lives on one of 
these branes, it is non--minimally coupled to the moduli. Without any  
stabilization mechanism for the moduli, the inflaton acquires a non--canonical 
kinetic term and, consequently, its mass varies with time. Therefore, for a given 
potential the presence of the moduli can alter two predictions of
inflation from standard General Relativity. Firstly, the duration of the inflationary 
period may change because the slope of the potential changes. Secondly, 
the evolution of perturbations will change 
because of the presence of other fields. Potentially, entropy perturbations 
will be generated and the slope of the power spectra will also be 
affected. In this paper we will investigate two popular cases  
from which we can learn about the influence of brane world moduli on the 
dynamics of inflation and the resulting perturbations. 
The first case is the quadratic potential, $V(\chi) = \frac{1}{2} m^2 \chi^2$, and the second 
is the quartic potential, $V(\chi) = \lambda \chi^4$. Considering 
inflation with these potentials in the context of General Relativity
one generates  the
following results, [\ref{linde}, \ref{liddlelyth}]:
\begin{itemize} 
\item For the quadratic potential one expects the spectral index of the 
curvature perturbation to be $n_R \approx 0.96$ and the spectral index of 
the tensor perturbations $n_T \approx -0.02$. 
\item In contrast, the  quartic potential generates a  curvature 
perturbation with  $n_R \approx 0.94$ and and gravitational waves with
$n_T \approx -0.04$.
\end{itemize}
As we will show in this paper, these predictions can change significantly if
moduli fields with large enough couplings are present during the inflationary 
stage. 

We will also discuss another, more problematic feature of 
the kind of brane world theories we consider: the second brane collapses 
during inflation driven by a scalar field on the positive tension brane. 
Such a behavior was found in [\ref{Brax:2002nt}] during matter domination, 
but as we will see, such behavior will be found during
inflation as well. It suggests that the five--dimensional spacetime is not 
stable during inflation (and the matter dominated epoch). 

The paper is organized as follows: in the next Section we will discuss
the moduli space approximation and state the conditions for this approximation 
to be valid. In Section 3 we will discuss the case in which the moduli
fields acquire a potential due to detuning of the brane tension. We show 
that in this case the potential is too steep to support a period 
of slow--roll inflation. In Section 4 we will study the case of an inflaton 
field on the positive tension brane and study in detail the cases of the 
quadratic and the quartic potential. Finally, we present our conclusions in Section 5.

\section{The Four--Dimensional Effective Action}
We consider a five--dimensional brane world setup with two boundary branes 
and a bulk scalar field $\psi$. The bulk scalar field induces a tension 
$|U_B| = 4k e^{\alpha \psi}$ with opposite signs on each branes. In addition, 
there is a bulk potential energy $U(\psi)$ for the bulk scalar which is related 
to $U_B(\psi)$ by
\begin{equation}\label{bpscondition}
U(\psi) = \left(\frac{\partial U_B(\psi)}{\partial\psi}\right)^2 - U_B(\psi)^2.
\end{equation}
The higher--dimensional theory contains the following 
parameters: the five--dimensional gravitational coupling 
constant $\kappa_5$, the energy scale $k$ of $U_B$ and $\alpha$. One can show 
[\ref{Brax:2002nt}] that $\alpha$ determines how warped the extra dimension is: large 
values of $\alpha$ corresponds to slightly warped bulk geometries, whereas small 
$\alpha$ corresponds to highly warped geometries. In fact, the case $\alpha=0$ 
is equivalent  to the Randall--Sundrum two--brane scenario, in which the 
spacetime between the branes is a slice of an five--dimensional Anti--de Sitter
spacetime. In this paper, we are interested in the limit in which a four--dimensional 
description is valid. The four--dimensional effective action can be 
constructed via the moduli space approximation, described in [\ref{Brax:2002nt}].
The reader is referred to [\ref{Brax:2002nt}] for details of the calculations; here
we will only summarize the main results and the condition for the moduli space 
approximation to hold. 

In the Einstein frame, the gravitational sector of the effective 
four--dimensional action is given by
\begin{eqnarray}
\mathcal{S}_{\rm EF} = \frac{1}{16\pi G} \int
d^4x \sqrt{-g} \left[ \mathcal{R} - \frac{12\alpha^2}{1+2\alpha^2}
(\partial \Phi)^2 - \frac{6}{1 + 2\alpha^2} (\partial R)^2\right],
\end{eqnarray}
where the gravitational constant $G$ is related to $\kappa_5$, $\alpha$ and 
$k$ by
\begin{eqnarray}
16\pi G = 2\kappa_5^2 (1 + 2 \alpha^2) k.
\end{eqnarray}
Therefore, given the four--dimensional gravitational constant $G$, there are only 
two free parameters in the theory. 

Because the bulk potential energy is directly related to the brane tensions, 
equation (\ref{bpscondition}), there is no contribution from the bulk scalar field 
to the cosmological constant. However, it is possible to introduce matter as 
well as potentials on the each of the 
branes. Let us consider first two potentials, $V(\Phi, R)$ and
$W(\Phi, R)$ which are to be seen as small, on the 
positive and negative tension brane respectively. Then, after transforming 
into the Einstein frame one finds [\ref{Brax:2002nt}]:
\begin{eqnarray}
V_{\rm eff} = A^4(\Phi,R) V(\Phi,R),\\
W_{\rm eff} = B^4(\Phi,R) W(\Phi,R),
\end{eqnarray}
with
\begin{eqnarray}
A^2(\Phi,R) &=& e^{-\frac{4\alpha^2}{1+2\alpha^2}\Phi} (\cosh R )^{\frac{2}{1+2\alpha^2}},\label{eq:conf1}\\
B^2(\Phi,R) &=& e^{-\frac{4\alpha^2}{1+2\alpha^2}\Phi} (\sinh R )^{\frac{2}{1+2\alpha^2}}.
\end{eqnarray}

The action for matter confined on each branes is
\begin{eqnarray}
\mathcal{S}_m^{(1)} ( \Psi_1, g_{\mu\nu}^{B_1} ) + \mathcal{S}_m^{(2)}
( \Psi_2, g_{\mu\nu}^{B_2} ).
\end{eqnarray}
We denote the matter fields on each brane by $\Psi_{1,2}$ and the induced 
metrics on each branes by $g_{\mu\nu}^{B_1,B_2}$ respectively. 
Transforming the matter action to the Einstein frame results
in the following full action:
\begin{eqnarray}
\mathcal{S}_{\rm EF} &=&\frac{1}{16\pi G} \int
d^4x \sqrt{-g} \left[ \mathcal{R} - \frac{12\alpha^2}{1+2\alpha^2}
\left(\partial \Phi \right)^2 - \frac{6}{2\alpha^2 + 1} (\partial
R)^2\right] \nonumber \\
&& \hspace{0.5cm} - \int d^4x \sqrt{-g} \left[ V_{\rm eff}(\Phi,R) + W_{\rm eff}(\Phi,R)
\right] \nonumber \\
&& \hspace{1cm} + \mathcal{S}_m^{(1)} ( \Psi_1, A^2(\Phi,R) g_{\mu\nu} ) 
+ \mathcal{S}_m^{(2)}
( \Psi_2, B^2(\Phi,R) g_{\mu\nu} ).\label{eq:mattercoup}
\end{eqnarray}
The origin of the potentials $V_{\rm eff}(\Phi,R)$ and $W_{\rm
  eff}(\Phi,R)$ might be a  supersymmetry breaking processes [\ref{Brax:2002nt}] 
(i.e. detuning the brane tensions) 
or they could be generated by quantum processes [\ref{garriga}]. The functions 
$A(\Phi,R)$ and $B(\Phi,R)$ determine the coupling of matter to the 
moduli fields and are different functions. Indeed, in the Einstein frame, 
the energy conservation equation takes the following form
\begin{eqnarray}
D_\mu T^{\mu\nu}_i =  \alpha_\Phi^i (\partial^\nu \Phi) T_i +  \alpha_R^i
( \partial^\nu R) T_i,
\end{eqnarray}
with
\begin{eqnarray}
\begin{array}{cccccc}
\alpha_\Phi^{(1)} &=& \frac{\partial \ln A}{\partial \Phi},  &     \alpha_\Phi^{(2)} &=& \frac{\partial \ln B}{\partial \Phi},\\ 
\alpha_R^{(1)} &=& \frac{\partial \ln A}{\partial R},  &     \alpha_R^{(2)} &=& \frac{\partial \ln B}{\partial R}.
\end{array}
\end{eqnarray}

The functions $A$ and $B$ are different because the higher dimensional 
spacetime is generally warped. In the extreme case $\alpha =0$, in which the bulk is 
a slice of an Anti--de Sitter spacetime, $A = \cosh R$ and $B = \sinh R$, whereas in 
the other extreme case for very large $\alpha$ one obtains $A=B\sim \exp(-\Phi)$. 
The latter case corresponds to only slightly warped bulk geometries so that the induced 
metrics on each branes coincide. 

The effective four--dimensional action (\ref{eq:mattercoup}) is a bi--scalar 
tensor theory and therefore subject to constraints. 
If the fields $\Phi$ and $R$ are strictly massless, observations constrain the 
coupling functions to be small, which implies that (see
[\ref{Brax:2002nt}] and [\ref{Damour:1992we}]):
\begin{eqnarray}
\alpha \lesssim 10^{-2}, \ \ R \lesssim 0.2. \label{eq:paramconst}
\end{eqnarray}
The value of $R$ must hold today but $R$ could be larger in the early universe. 
However, if  $R$ is initially large, the predictions for the CMB are affected 
[\ref{Rhodes:2003ev}].

When is the four--dimensional action (\ref{eq:mattercoup}) a good description 
for the dynamics of the system? In deriving the action (\ref{eq:mattercoup}), 
it was assumed that the fields $\Phi$ and $R$ evolve slowly so that 
higher derivatives can be neglected. Furthermore, 
any additional matter on the branes with energy density $\rho$ should contribute only 
slightly to the total energy density of the branes $\rho_{\rm tot} = U_B + \rho$, 
so that $\rho \ll U_B$. In practical terms, we consider scales below the brane 
tension and therefore a regime in which quadratic corrections to 
Einstein's equation are negligible. Any heavy Kaluza--Klein modes are assumed to 
be negligible and are integrated out. If inflation takes place in the 
high--energy regime, massive modes are likely to be excited. Furthermore, 
we have to keep in mind that length scales which are of cosmological interest 
today might have their origin in scales which are much shorter than the 
higher--dimensional Planck scale or the bulk curvature scale. It is not 
clear {\it a priori} that the use of the action (\ref{eq:mattercoup}) is fully 
justified. Because of this, the transplanckian problem of 
inflationary cosmology [\ref{transpa}] is even more acute in brane cosmology. 
The aim of this paper, however, is to show that even if one neglects 
transplanckian physics, there are new effects in brane cosmology due to 
the existence of two branes and a bulk scalar field. 

In the case of brane cosmology (and, in fact, in scalar--tensor theories in 
general), there is an additional reason to study a phase of inflation, apart 
from the usual problems of the standard cosmology. Namely, we find that 
an attractor mechanism operates during inflation, in which the coupling of the moduli fields 
to matter is driven towards small values and thereby makes the theory more 
viable when compared to observations (similar to the ones discussed in [\ref{wands}]
--[\ref{damour}). However, we will see that this attractor in our theory implies 
that the higher dimensional spacetime is unstable during inflation.

\section{Inflation Driven by the Moduli and Exponential Potentials}
Before we discuss our primary objective, inflation 
driven by a scalar field confined on the positive tension brane, in this 
section we entertain the possibility that inflation might be driven by 
the moduli fields themselves. The potential energy for the fields is obtained 
by detuning the positive brane tension away from its critical value so that 
a four--dimensional potential energy appears. Again, we shall consider 
inflation on the positive tension brane only, as this is the one we must live on
if the moduli fields are not stabilized. The resulting potential is then given 
by [\ref{brax:2000xk}] 
\begin{eqnarray}
V = (T-1) 4k e^{\alpha \psi},
\end{eqnarray}
where $T \neq 1$ is the supersymmetry breaking parameter. Now writing
this in terms of $\Phi$ and $R$ and setting $4(T-1)k = V_0$ one generates 
[\ref{Brax:2002nt}],
\begin{eqnarray}
V_{\rm eff}(\Phi, R) = V_0 \exp\left[ - \frac{12\alpha^2 }{1+2\alpha^2}\Phi
\right] \left( \cosh R \right)^{\frac{4-4\alpha^2}{1+2\alpha^2}}.\label{eq:modpotential}
\end{eqnarray}
The equations of motion resulting from the effective action (\ref{eq:mattercoup})
are then given by 
\begin{eqnarray}
H^2 &=& \frac{8\pi G}{3} V_{\rm eff} + \frac{2\alpha^2}{1+2\alpha^2}
\dot{\Phi}^2 + \frac{1}{1+2\alpha^2} \dot{R}^2,\\
\dot{H} &=& -\frac{6\alpha^2}{1+2\alpha^2} \dot{\Phi}^2 
- \frac{3}{1+2\alpha^2} \dot{R}^2,\\
\ddot{R} + 3H \dot{R} &=& -8\pi G \frac{1 + 2\alpha^2}{6} \frac{\partial
V_{\rm eff}}{\partial R},\\
\ddot{\Phi} + 3H \dot{\Phi} &=& -8\pi G \frac{1+2\alpha^2}{12\alpha^2}
\frac{\partial V_{\rm eff}}{\partial \Phi}.
\end{eqnarray}

We notice that the potential we get has many similar features to those
studied in [\ref{Ashcroft:2002vj}]. The reader is reminded that 
$\alpha \ll 1$, see (\ref{eq:paramconst}), to give compliance with tests
on the equivalence principle.  A quick glance suggests that $R$ may
fast roll whereas $\Phi$ may be a candidate for slow--roll
inflation. If this were to happen with minimal evolution of $\Phi$, then $R$ would quickly settle in
the minimum and $\Phi$ alone would support a period of power law
inflation. This would look like a single field, therefore, it would be interesting if both
fields were important dynamically for inflation. Here we are
able to study this by assuming nothing about slow--roll. Let us approximate the potential by
\begin{eqnarray}
V_{\rm eff} (\Phi, R) \approx  \frac{V_0}{2} \exp\left[ - \frac{12\alpha^2 }{1+2\alpha^2}\Phi
\right] \exp \left[ \frac{4-4\alpha^2}{1+2\alpha^2}R \right],
\end{eqnarray}
and see what happens for large positive $R$. Having written the potential like this, we immediately see 
\begin{eqnarray}
\frac{\partial V_{\rm eff}}{\partial R} &=& \frac{
 4-4\alpha^2}{1+2\alpha^2} V_{\rm eff},\\
\frac{\partial V_{\rm eff}}{\partial \Phi} &=& - \frac{12\alpha^2
 }{1+2\alpha^2} V_{\rm eff}.
\end{eqnarray}
This allows us to analyze the dynamical system and study the critical
 points in the three--dimensional phase space. Let us
transform to the dimensionless parameters
\begin{eqnarray}
x &=& \left(\frac{2\alpha^2}{1+2\alpha^2}\right)^{\frac{1}{2}}\frac{\dot{\Phi}}{H},\\
y &=& \left(\frac{1}{1+2\alpha^2}\right)^{\frac{1}{2}}\frac{\dot{R}}{H},\\
z &=& \left(\frac{3}{8\pi G}\right)^{\frac{1}{2}} \frac{\sqrt{V_{\rm eff}}}{H}.
\end{eqnarray}
Substituting this into our equations of motion we find that
\begin{eqnarray}
x' &=& 3(x^2 + y^2) x + 3z^2 \left(\frac{2\alpha^2}{1+2\alpha^2}
\right)^{\frac{1}{2}} - 3x,\\
y' &=& 3(x^2 + y^2) y - z^2 \frac{2-2\alpha^2}{( 1+ 2\alpha^2)^{\frac{1}{2}}}
-3y,\\
z' &=& 3(x^2 + y^2) z + z \left[ \frac{2 -
2\alpha^2}{(1+2\alpha^2)^{\frac{1}{2}}} y - 3\left( \frac{2\alpha^2
}{1+2\alpha^2} \right)^{\frac{1}{2}} x \right],\\
1 &=& x^2 + y^2 + z^2,
\end{eqnarray}
where $' \equiv d/d\ln a$, $a(t)$ the usual scalefactor. Writing
\begin{eqnarray}
\lambda &=&
3\left(\frac{2\alpha^2}{1+2\alpha^2}\right)^{\frac{1}{2}},\\
\mu &=& \frac{2 - 2\alpha^2}{(1+ 2\alpha^2)^{\frac{1}{2}}},
\end{eqnarray}
it is easy to find the critical points of the dynamical system and to
study their stability. The results are summarized in Table
\ref{tab:crit} where we write the point as $(X,Y,Z)$.
\begin{table}[!ht]
\caption{The properties of the critical points for the coupled
potential of the moduli fields}\label{tab:crit}
\center{
\begin{tabular}{ccc|c}
\hline
\hline
$X$ & $Y$ & $Z$ & Stability   \\ 
\hline
1 & 0  & 0 & Unstable Node for $ \lambda < 3$  \\ 
 & & & Saddle point for $ \lambda > 3$  \\
-1 & 0  &0 & Unstable Node for $ \lambda >
- 3$  \\
&&& Saddle point for $ \lambda <
-3$  \\
\hline
0 & 1 & 0 & Unstable Node for$ \mu <
3$  \\ 
&&& Saddle point for $ \mu >
3$  \\
0 & -1 & 0 & Unstable Node for $ \mu >
-3$  \\
&&& Saddle point for $ \mu <
-3$  \\
\hline
$\gamma$ &  $(1-\gamma^2)^{1/2}$ & 0&
$0\leq\gamma\leq 1 $  Saddle  \\ 
\hline
 $\frac{\lambda}{3}$ &$- \frac{\mu}{3} $&$\left(1- \frac{\lambda^2}{9} -
\frac{\mu^2}{9}\right)^{1/2}$ & Stable point with  Hubble constraint \\
\hline
\hline
\end{tabular}
}
\end{table}
We see that the system has one stable critical point only. The
question is whether this gives us an epoch of slow--roll inflation. 
We can test this simply by calculating
\begin{eqnarray}
\epsilon &=& - \frac{\dot{H}}{H^2},\\
&=& 3(X^2 + Y^2). 
\end{eqnarray}
It is easy to show that this gives us the condition for inflation
\begin{eqnarray}
\lambda^2 + \mu^2 \ll 3.
\end{eqnarray}
This then generates the constraint on $\alpha$,
\begin{eqnarray}
(2\alpha^2 + 1 )^2 \ll 0.
\end{eqnarray}
This, of course, is impossible to satisfy for any real $\alpha$ and so
it appears that the two moduli fields are unable to support a period 
of slow--roll inflation where both are important dynamically. Of course, the
approximation made in the potential will break down as $R$ becomes
small and we generate inflation through $\Phi$ as mentioned earlier. 

To summarize, if $R$ is initially large and the brane 
tensions are detuned away from their critical value, the resulting 
potential is too steep to support a period of slow--roll inflation. 
When $R$ is small it does not contribute to the dynamics 
of inflation, and the resulting inflationary period is of power--law type 
driven by $\Phi$. Of course, one might consider other potentials for 
the fields $R$ and $\Phi$, in which case the conclusions drawn above 
will not hold. However, the origin of such potentials for the moduli 
fields are not yet clear. Furthermore, even if the moduli obtain 
a potential, for example due to quantum effects [\ref{garriga}], it is not clear that 
the resulting potential will have the properties to drive a period of 
inflation, in which the resulting perturbation spectra are compatible 
with observations.

\section{Boundary Inflation driven by Matter on the Branes}
We now study the case of inflation driven by an inflaton on the positive 
tension brane, dubbed boundary inflation [\ref{kobayashi}, \ref{lukas}]. In general, 
there will be new effects when inflation is driven by a scalar field 
confined to one of the branes. In particular, if the energy scale
of the inflaton field is larger than the brane tension the 
five--dimensional theory has to be used. Potentially, Kaluza--Klein 
modes might be excited and affect the predictions for scalar and 
tensor perturbations. However, we do not include those modes here. 
Instead, we concentrate on the zero mode of the graviton and on 
energy scales less than the brane tension so that the effective 
action presented in Section 2 should be a good description of the 
dynamics of the system. We point out that our setup is different from the 
one discussed in [\ref{kofman}]. It is clear from the discussion in
Section 2 that 
the moduli fields $R$ and $\Phi$ couple non--minimally to the 
inflation field. This will result in new effects for both the 
background and perturbations.

\subsection{Background evolution}
Let us denote the inflaton on the positive tension brane by $\chi$. From 
(\ref{eq:mattercoup}), we see that, for $V(\chi) = V_0 \chi^n$, we generate 
a matter action
\begin{eqnarray}
\mathcal{S}_{\rm Matter} = \int \sqrt{-g} d^4x \left[ -\frac{1}{2}
A^2(\Phi, R) (\partial \chi)^2 - A^4(\Phi,R) V_0 \chi^n
\right],
\end{eqnarray}
with $A(\Phi, R)$ given by 
(\ref{eq:conf1}). Since we assume there is no potential for the
moduli fields, the resulting Einstein frame action 
is equivalent to one with three dynamical fields all coupled
through a single potential. Let us write this single potential as
\begin{eqnarray}
\tilde{V}(\Phi, R , \chi ) &=& A^4(\Phi, R) V_0 \chi^n,\\
&=& V_0 \exp\left[-\frac{8\alpha^2}{1+2\alpha^2}\Phi\right] (\cosh R
)^{\frac{4}{1+2\alpha^2}} \chi^n.
\end{eqnarray}
Then we find that the equations of motion are given
by
\begin{eqnarray}
H^2 &=& \frac{2\alpha^2}{1+2\alpha^2} \dot{\Phi}^2 +
\frac{1}{1+2\alpha^2} \dot{R}^2\nonumber \\
&&\hspace{0.5cm} + \frac{4\pi G}{3} \left[ A^2(\Phi,R)
\dot{\chi}^2 + 2 \tilde{V}(\Phi, R, \chi)\right],\\
\ddot{\Phi} + 3H \dot{\Phi} &=& 8\pi G \frac{1+2\alpha^2}{12\alpha^2}
\left[ A(\Phi,R) \frac{\partial A}{\partial \Phi} \dot{\chi}^2 -
\frac{\partial \tilde{V}}{\partial \Phi} \right],\\
\ddot{R} + 3H \dot{R} &=& 8\pi G \frac{1+2\alpha^2}{6}
\left[ A(\Phi,R) \frac{\partial A}{\partial R} \dot{\chi}^2 -
\frac{\partial \tilde{V}}{\partial R} \right],\\
\ddot{\chi} + 3H \dot{\chi} &=& -\frac{2}{A(\Phi,R)}\dot{\chi}
\left[ \dot{\Phi} \frac{\partial A}{\partial \Phi}  +
\dot{R} \frac{\partial A}{\partial R} \right] -
\frac{1}{A^2(\Phi,R)} \frac{\partial \tilde{V}}{\partial \chi}.
\end{eqnarray}

Let us now make the further assumptions that $R$ fast--rolls down to
its minimum $R=0$ and $\Phi$ and $\chi$ slow--roll. This seems a reasonable
approximation to make at this stage because $\alpha \ll 1$. Let us
also assume that $R$ rolls sufficiently quickly for the other fields 
not to evolve. If $R$ is initially very large, it dominates the potential 
energy and rolls quickly towards zero. If it is small initially, its coupling 
to the inflaton is small and it does not contribute to the total energy density. 
In this case, perturbations in $R$ can be neglected. 
Of course it could play an important role after inflation which we 
do not discuss in this paper. However, the fact that $R$ quickly approaches 
zero has very important implications for model building. We will return to this 
point later. 

With these assumptions, the system simplifies significantly and we are
left with the equations of motion
\begin{eqnarray}
H^2 &=& \frac{8\pi G}{3} \tilde{V}(\Phi, 0 , \chi), \\
3H\dot{\Phi} &=& - 8\pi G \frac{1+2\alpha^2}{12\alpha^2}
\frac{\partial \tilde{V}}{\partial \Phi}(\Phi, 0 , \chi),\label{phieq}\\
3H\dot{\chi} &=& - \frac{1}{A^2(\Phi, 0 )}
\frac{\partial \tilde{V}}{\partial \chi}(\Phi, 0 , \chi).
\end{eqnarray}

It is straightforward to derive the relationship between the two fields, 
\begin{eqnarray}
\Phi = \frac{1+ 2\alpha^2}{4\alpha^2} \ln \left[
\frac{4\alpha^2}{1+2\alpha^2} \left( k_1 - 8\pi G \frac{\chi^2}{3n}
\right)\right],
\end{eqnarray}
where $k_1$ is an integration constant determined by the initial conditions.
Note that this relation (and the following solutions) holds  strictly
for $\alpha \neq 0$ {\it only}, because $\Phi$ decouples for $\alpha = 0$ 
and equation (\ref{phieq}) becomes redundant. 

It is not too difficult to find a solution for the two fields
\begin{eqnarray}
\chi(t) &=& \left( k_2 - \frac{(4-n)n}{2} \left( \frac{8\pi G V_0}{3}
\right)^{\frac{1}{2}} t \right)^{\frac{2}{4-n}}, \\
\Phi(t) &=& \frac{1+2\alpha^2}{4\alpha^2} \ln
\left[\frac{4\alpha^2}{1+2\alpha^2} \right. \nonumber \\
&&\times \left. \left( k_1 - \frac{8\pi G}{3n} \left(
k_2 - \frac{(4-n)n}{2}\left( \frac{8\pi G V_0}{3}
\right)^{\frac{1}{2}} t \right)^{\frac{4}{4-n}} \right)\right],
\end{eqnarray}
for the case $n\neq4$. We shall cover this special example in a moment. 
We can evaluate the integration constant $k_2$ from the initial condition 
for $\chi$. We then find that the Hubble parameter behaves as
\begin{eqnarray}
H(t) &=& \left( \frac{8\pi G V_0}{3}
\right)^{\frac{1}{2}}\frac{1+2\alpha^2}{4\alpha^2} \left[ k_2 
- \frac{(4-n)n}{2}\left( \frac{8\pi G V_0}{3}
\right)^{\frac{1}{2}} t \right]^{\frac{n}{4-n}} \nonumber \\
&& \times \frac{1}{ \left( k_1 - \frac{8\pi G}{3n} \left(
k_2 - \frac{(4-n)n}{2}\left( \frac{8\pi G V_0}{3}
\right)^{\frac{1}{2}} t \right)^{\frac{4}{4-n}} \right)     }.
\end{eqnarray}
This then gives
\begin{eqnarray}
a(t) &\propto&  \left( k_1 - \frac{8\pi G}{3n} \left(
k_2 - \frac{(4-n)n}{2}\left( \frac{8\pi G V_0}{3}
\right)^{\frac{1}{2}} t \right)^{\frac{4}{4-n}}
\right)^{\frac{3(1+2\alpha^2)}{8\alpha^2
}}  .
\end{eqnarray}
  It should be added that when $n>0$ we require
the inflaton to roll down the potential to its minimum. Before it
reaches this point it ceases to slow--roll and the solutions are no
longer valid. This prevents us from generating negative
quantities.

Now let us turn our attention to the case of $n=4$. As we have said already
 this turns out to be soluble. The results are stated below.
\begin{eqnarray}
\chi(t) &=& \chi_0 \exp\left[ -4 \left( \frac{8\pi G V_0}{3}
\right)^{\frac{1}{2}} t \right],\\
\Phi(t) &=& \frac{1+ 2\alpha^2}{4\alpha^2} \ln \left[
\frac{4\alpha^2}{1+2\alpha^2} \left( k_1 - 8\pi G\frac{\chi_0^2}{12}\exp\left[ -8 \left( \frac{8\pi G V_0}{3}
\right)^{\frac{1}{2}} t \right]
\right)\right].\\
H(t) &=& \left( \frac{8 \pi G V_0}{3} \right)^{\frac{1}{2}} \chi_0^2
\frac{1+2\alpha^2}{4\alpha^2} \frac{ \exp\left[ -8 \left( \frac{8\pi G V_0}{3}
\right)^{\frac{1}{2}} t \right]}{\left( k_1 - 8\pi G\frac{\chi_0^2}{12}\exp\left[ -8 \left( \frac{8\pi G V_0}{3}
\right)^{\frac{1}{2}} t \right]
\right)},\\
a(t) &\propto& \left(k_1 - 8\pi G\frac{\chi_0^2}{12}\exp\left[ -8 \left( \frac{8\pi G V_0}{3}
\right)^{\frac{1}{2}} t
\right]\right)^{\frac{3(1+2\alpha^2)}{8\alpha^2}}. \label{eq:branefactor1}
\end{eqnarray}
We should point out that for large enough time
it appears as though $a(t) \rightarrow \mathrm{constant}$ in (\ref{eq:branefactor1}). This is an artifact of the slow--roll approximation and we
would expect inflation to end well before this would occur. After this
other dynamical considerations would proliferate.

Numerical simulations verify these solutions. The problem of
initial conditions still remains however. Furthermore, we are able to
obtain an approximate solution for the $R$ field. As predicted, this
rapidly rolls down to the minimum of its potential. In addition, it
undergoes no oscillations as one may expect for a generic symmetric
potential with a minimum of this type. Such oscillations about zero
would correspond to the brane oscillating about the singularity which
would become apparent to an observer on the brane. This would  not be a
desirable feature. 

\begin{figure}[!ht]
\centerline{\includegraphics{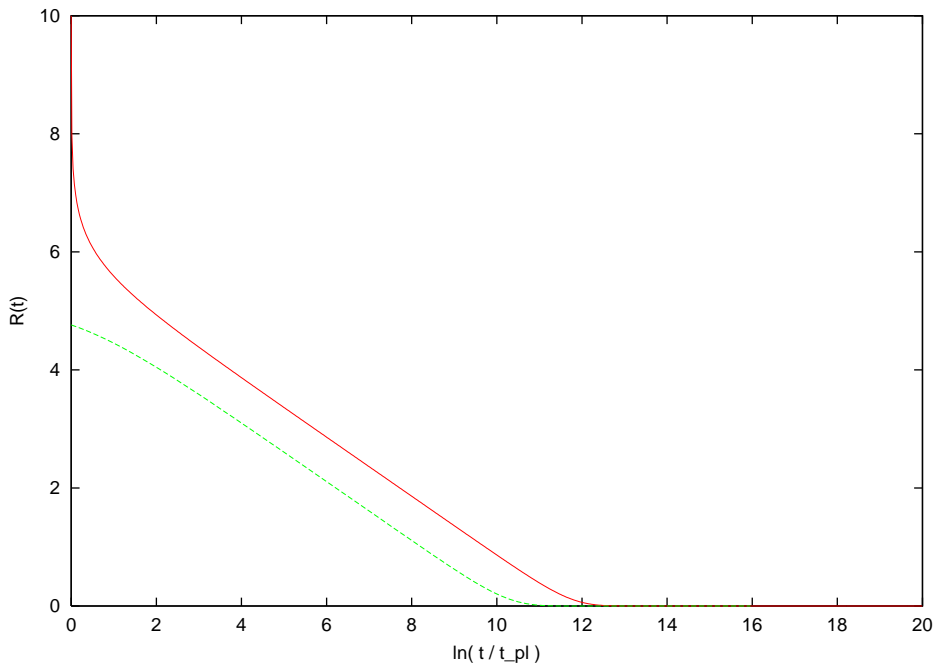}}
\caption{ The evolution of $R$ is plotted for the analytic solution,
  equation (\ref{eq:rsoln}),  with
  dashed line (green) 
and the numerical one with a solid  line (red).}\label{fig:revolve}
\end{figure}

Since the form of the potential looks approximately like an
exponential $V(R) \sim \exp[4R]$, we see that the slow--roll
parameters are of $\mathcal{O}(1)$. Thus we would expect the
slow--roll solution to be a reasonable approximation. Furthermore we
neglect $\alpha^2$ because we must have $\alpha < 0.1$. With this
in mind, one is able to show that 
\begin{eqnarray}
\tanh R = \kappa \exp\left[ - 4\left( \frac{I_0}{3}
\right)^{\frac{1}{2}} t \right],
\end{eqnarray}
where $\kappa$ is an integration constant and $I_0 = V_0 \chi_0^n
\exp\left[ \frac{-8\alpha^2}{1+2\alpha^2} \Phi_0 \right]$. We have
used the initial values of $\chi, \Phi$ in the potential and we
assume that we do not deviate from these during the evolution of
$R$. Now since $\kappa$ depends upon $\tanh R_0$, one
sees that it is weakly related to the initial condition. We shall take
$\kappa =1$ as the generic case. Then it is easy to show that the
solution is given by
\begin{eqnarray}
R = \frac{1}{2} \ln \left[ \frac{1 + \exp\left[ - 4\left( \frac{I_0}{3}
\right)^{\frac{1}{2}} t \right]  }{ 1 - \exp\left[ - 4\left( \frac{I_0}{3}
\right)^{\frac{1}{2}} t \right]  }\right].\label{eq:rsoln}
\end{eqnarray}
It is also clear from this that $R$ decays quickly but approaches the
singularity exponentially. The behavior of the analytic solution against 
the numerical one is shown in Figure \ref{fig:revolve}.

We would like to point out that the point 
$R$ is special and problematic: if $R=0$, the negative tension brane 
sits at a singularity in the fifth dimension [\ref{Brax:2002nt}]. 
Thus, while $R$ approaches zero, the negative tension brane collapses. 
Although during inflation the negative tension brane never reaches the 
singularity, the collapse of the second brane means that the spacetime 
is unstable. A similar behavior was found in the matter dominated 
era in [\ref{Brax:2002nt}]. To avoid the collapse of the second brane, 
one can either consider a potential which stabilizes $R$ at a finite 
value $R>0$ or keeps the second brane away from the singularity. 
Alternatively, and more speculatively, one can consider matter forms 
on the second brane which violate the strong energy condition. We 
do not consider these alternatives in this paper. Instead, we will 
assume that $R$ is small and does not contribute to the dynamics for
both background and perturbations. 

This concludes the discussion on the background evolution. 
There is the possibility of extending the considerations by 
studying other potentials. Some qualitative aspects of these 
are covered in part in [\ref{Ashcroft:2002vj}].

\subsection{Perturbation evolution during inflation}

We turn now our attention to the consequences for cosmological 
perturbations generated during inflation. We will use the formalism 
discussed in [\ref{DiMarco:2002eb}], which is an extension of the
formalism presented in [\ref{Gordon:2000hv}]. We remind the reader 
again that we assume that $R$ does not play a role at least
in the last 60 e--folds of inflation.

\subsubsection{Perturbation equations}

The equations of motion for the fields $\Phi$ and $\chi$ are given by
\begin{eqnarray}
\Box \Phi &=& \frac{1+2\alpha^2}{12\alpha^2} \left[ A \frac{\partial
A}{\partial \Phi} g^{\mu\nu} \partial_\mu \chi \partial_\nu \chi +
\frac{\partial \tilde{V}}{\partial \Phi} \right].\\
\nabla_\mu \left[ A^2 g^{\mu\nu} \partial_\nu \chi \right] &=&
\frac{\partial \tilde{V}}{\partial \chi}.
\end{eqnarray}

We now perturb the two fields
\begin{eqnarray}
\Phi &\rightarrow& \Phi(t) + \delta\Phi(\vect{x},t),\\
\chi&\rightarrow& \chi(t) + \delta\chi(\vect{x},t).
\end{eqnarray}
Furthermore, we choose to work in the longitudinal gauge so that the perturbed
metric looks like
\begin{eqnarray}
ds^2 = -(1 + 2 \Psi ) dt^2 + a^2 ( 1-2\Psi )d\vect{x}^2.
\end{eqnarray} 
This generates the following equations of motion:
\begin{eqnarray}
\delta\ddot{\Phi} + 3 H \delta \dot{\Phi} + \frac{k^2}{a^2}\delta \Phi &=&
\frac{1+2\alpha^2}{12\alpha^2} \left[ \left(\left( \frac{\partial A}{\partial
\Phi} \right)^2 \dot{\chi}^2  + A \frac{\partial^2
A}{\partial \Phi^2} \dot{\chi}^2\right) \delta \Phi \right.\nonumber \\
&& \left.+ 2  A \frac{\partial A}{\partial \Phi}\dot{\chi}
\delta\dot{\chi} - \frac{\partial^2
\tilde{V}}{\partial \Phi^2} \delta\Phi - \frac{\partial^2
\tilde{V}}{\partial \Phi\partial \chi} \delta\chi \right]\nonumber \\
&& -  \frac{1+2\alpha^2}{6 \alpha^2} \frac{\partial
\tilde{V}}{\partial \Phi}\Psi + 4 \dot{\Phi} \dot{\Psi}.\\
A^2\left[\delta\ddot{\chi} + 3 H \delta \dot{\chi} +
\frac{k^2}{a^2}\delta \chi\right] &=& \left[ \frac{2}{A}
\frac{\partial A}{\partial \Phi} \frac{\partial \tilde{V}}{\partial
\chi} + 2\left( \frac{\partial A}{\partial
\Phi} \right)^2 \dot{\chi}\dot{\Phi} - 2 A \frac{\partial^2
A}{\partial \Phi^2} \dot{\chi}\dot{\Phi}\right]\delta\Phi \nonumber \\
&& -  \frac{\partial^2
\tilde{V}}{\partial \chi^2} \delta\chi - \frac{\partial^2
\tilde{V}}{\partial \chi\partial \Phi} \delta\Phi\nonumber \\
&& - 2 A \frac{\partial A}{\partial \Phi}\left[ \dot{\chi} \delta\dot{\Phi}
+ \dot{\Phi} \delta\dot{\chi} \right] \nonumber\\
&& + 4A^2 \dot{\chi} \dot{\Psi} - 2 \frac{\partial \tilde{V}}{\partial
\chi} \Psi.
\end{eqnarray}
This set up is equivalent to Di Marco \emph{et al}
[\ref{DiMarco:2002eb}] provided we make
the field re--definition
\begin{eqnarray}
 \left( \frac{12 \alpha^2}{1 + 2\alpha^2} \right)^{\frac{1}{2}} \Phi =
 \Sigma.
\end{eqnarray}
so that, dropping the tilde,
\begin{eqnarray}
V(\Sigma, \chi ) &=& V_0 \chi^n \, e^{ 4 \beta \Sigma},\\
\beta &=&  \frac{-\alpha }{\sqrt{3}(1 + 2 \alpha^2)^{\frac{1}{2}}}
\end{eqnarray}
Once again note that $\alpha \rightarrow 0$ causes the modulus to
decouple. 
We now define two new fields, the adiabatic and entropy perturbations,
by
\begin{eqnarray}
\delta \sigma &=& \cos \theta \delta \Sigma + A \sin \theta \delta \chi,\\
\delta s &=& -  \sin \theta \delta \Sigma + A \cos \theta \delta \chi.
\end{eqnarray}
where
\begin{eqnarray}
\cos \theta &=& \frac{\dot{\Sigma}}{\sqrt{ \dot{\Sigma}^2 +
A^2\dot{\chi}^2}},\\
\sin \theta &=& \frac{A\dot{\chi}}{\sqrt{ \dot{\Sigma}^2 +
A^2\dot{\chi}^2}}.
\end{eqnarray}
Furthermore, the gauge invariant curvature and entropy perturbations $\mathcal{R}$ 
(denoted as $\zeta$ in [\ref{DiMarco:2002eb}]) and 
$\mathcal{S}$ are defined as 
\begin{eqnarray}
\mathcal{R} &=& \Psi + H \left( \frac{ \dot{\Sigma} \delta \Sigma
  + A^2\dot{\chi}\delta\chi}{   \dot{\Sigma}^2 +
A^2\dot{\chi}^2} \right) = \Psi+  H \frac{\delta \sigma}{\dot{\sigma}},\\
\mathcal{S} &=&  H \left(\frac{  A \dot{\Sigma} \delta \chi -A\dot{\chi}\delta\Sigma}{ \dot{\Sigma}^2 +
A^2\dot{\chi}^2} \right) = H \frac{\delta s}{\dot{\sigma}},
\end{eqnarray}
where $\dot{\sigma} = \sqrt{ \dot{\Sigma}^2 +
A^2\dot{\chi}^2}$. This demonstrates from where $\delta \sigma$ and $\delta
s$ derive their names. In addition, the equation of motion for $\sigma$
is given by,
\begin{eqnarray}
\ddot{\sigma} + 3 H \dot{\sigma} &=& -V_\sigma,\\
V_\sigma &=&  \cos \theta V_\Sigma + \frac{1}{A} \sin
\theta V_\chi.
\end{eqnarray}

We are now able to write the perturbation equations in terms of the 
gauge--invariant variables. They are given by [\ref{DiMarco:2002eb}]:
\begin{eqnarray}
\ddot{\mathcal{R}} &+& \left( 3H -2 \frac{\dot{H}}{H} +
  \frac{\ddot{H}}{\dot{H}} \right) \dot{\mathcal{R}} + \frac{k^2}{a^2} \mathcal{R} \\
  &=& \frac{H}{\dot{\sigma}} \left[ 2 ( \dot{\theta} \delta s )^. - 2 \left(
  \frac{V_\sigma}{\dot{\sigma}} + \frac{\dot{H}}{H} \right)
  \dot{\theta} \delta s +  2 \beta h(t) \right].\nonumber
\end{eqnarray}
\begin{eqnarray}
\ddot{\delta s} + 3H \dot{\delta s} + \left[\frac{k^2}{a^2}V_{ss} + 3\dot{\theta}^2  + \beta^2
  g(t)+ \beta f(t) - 4 \frac{V_s^2}{\dot{\sigma}^2} \right] \delta s = 2 \frac{V_s}{H} \dot{\mathcal{R}}.
\end{eqnarray}
where
\begin{eqnarray}
h(t) &=& \dot{\sigma}( \sin \theta \delta s )^. - \sin \theta \left[
  \frac{\dot{H}}{H} \dot{\sigma} + 2 V_\sigma - 3H\dot{\sigma} \right]
\delta s,\\
g(t) &=& - \dot{\sigma}^2 ( 1 + 3\sin^2 \theta ),\\
f(t) &=& V_\Sigma ( 1 + \sin^2 \theta ) - 4 V_s \sin \theta.
\end{eqnarray}
In addition we also define
\begin{eqnarray}
V_s &=& -  \sin \theta V_\Sigma + \frac{1}{A} \cos
\theta V_\chi,\\
V_{\sigma\sigma} &=&  \cos^2 \theta V_{\Sigma\Sigma} +
\frac{2}{A} \cos \theta \sin \theta V_{\Sigma\chi} +
\frac{1}{A^2} \sin^2 \theta V_{\chi\chi},\\
V_{\sigma s } &=& \left( \frac{1}{A^2} V_{\chi\chi} -
V_{\Sigma\Sigma} \right) \sin\theta  \cos\theta + \frac{1}{ A
}V_{\Sigma\chi}\left( \cos^2 \theta - \sin^2\theta \right),\\
V_{ss} &=& \sin^2 \theta V_{\Sigma\Sigma} -
\frac{2}{ A} \cos \theta \sin \theta V_{\Sigma\chi} +
\frac{1}{A^2} \cos^2 \theta V_{\chi\chi}.
\end{eqnarray}
   
One is then able to see the coupling between the entropy and curvature
perturbations. Note that whilst $V_{\sigma } = \frac{\partial
  V}{\partial \sigma}$ and $V_{s } = \frac{\partial
  V}{\partial s}$, it is not true that $V_{\sigma\sigma} =
\frac{\partial^2 V}{\partial \sigma^2}$ due to the dependence of the
non--canonical kinetic term on $\Sigma $. The same holds for $V_{\sigma s}$ and $V_{ss}$. Normally, without the non--canonical kinetic term, one
would find that there is no coupling if $\dot{\theta} = 0$. We now see
that, even if this holds, the coupling is maintained. Further, we see the
manner in which the perturbations source one and other. In addition
\begin{eqnarray}
\dot{\theta} = \dot{\sigma} \left[ - \frac{V_s}{\dot{\sigma^2}} -
  \beta\sin \theta \right],
\end{eqnarray} and so we see that the non--canonical kinetic term also
  introduces a source for $\theta$. This should be compared with the
  scenario in Randall--Sundrum where the entropy perturbations are
  generically suppressed due to the presence of the $\rho^2$ term in
  the Friedmann equation, [\ref{Ashcroft:2002ap}].
Also, note that the parameter $\beta$ which describes the source from the
  non--canonical kinetic term, is constant in this setup.
The next step is to examine the system numerically.

\subsubsection{Numerical Study of the Perturbation Spectra}
The method for the evolution of the perturbations is as follows: we
evolve the background staring with the initial conditions
\begin{eqnarray}
\Sigma_{init} &=& 0,\\
\chi_{init} &=& 1000,
\end{eqnarray}
for $n \geq 0$. This means that at the beginning of inflation the
inflaton mass will be equal to its physical one. When $-\dot{H}/H^2 = 1$, we stop the
evolution as this is where the scale factor ceases to accelerate. We then backtrack 67
e--folds and switch on the perturbations where each
mode starts inside the horizon with $k = 1000 aH$ at different
times. With this prescription, the largest modes undergo 60 e--folds
of inflation after horizon exit. We follow this evolution until the end of inflation.

To set the initial conditions one treats the modes, $\delta\sigma$
and $\delta s$, as independent stochastic variables deep inside 
the horizon. With this prescription, it can be shown that
\begin{eqnarray}
a\delta \sigma  =  \frac{1}{\sqrt{2k}} e^{-\imath k \tau},
\end{eqnarray}
deep inside the horizon where the universe looks like Minkowski space.
 To calculate the spectra numerically, we use the 
method described in [\ref{Tsujikawa:2002qx}]. 
One makes two runs: the first run begins in the Bunch--Davis vacuum for $\delta
\sigma$, $\delta s = \delta \dot{s} = 0$,
and the second in the Bunch--Davis vacuum for $\delta s$. 
The power spectra are then given by
\begin{eqnarray}
{\mathcal{P_R}} &=& \frac{k^3}{2\pi^2} \left( |{\mathcal{R}}_1|^2 +  |{\mathcal{R}}_2|^2
\right),\\
{\mathcal{P_S}} &=& \frac{k^3}{2\pi^2} \left( |{\mathcal{S}}_1|^2 +  |{\mathcal{S}}_2|^2
\right),\\
{\mathcal{P_C}} &=& \frac{k^3}{2\pi^2} | {\mathcal{R}}_1 {\mathcal{S}}_1 +
{\mathcal{R}}_2 {\mathcal{S}}_2|,
\end{eqnarray}
where the subscripts $1,2$ are the results for the two different runs. 
Furthermore, for the tensor modes,  one finds 
\begin{eqnarray}
P_{\mathcal{T}} = 8 \left( \frac{H_*}{2\pi} \right)^2,
\end{eqnarray} 
we also define
\begin{eqnarray}
r_{\mathcal{C}} = \frac{P_{\mathcal{C}}}{\sqrt{
      P_{\mathcal{R}}P_{\mathcal{S}} } }, \ \ \ \ \ r_{\mathcal{T}} =
      \frac{P_T}{16 P_{\mathcal{R}}}
\end{eqnarray}
Note that the definition for $r_\mathcal{T}$ is 16 times smaller than that used by the WMAP team, [\ref{Peiris:2003ff}].

\begin{figure}[!ht]
\centerline{\includegraphics{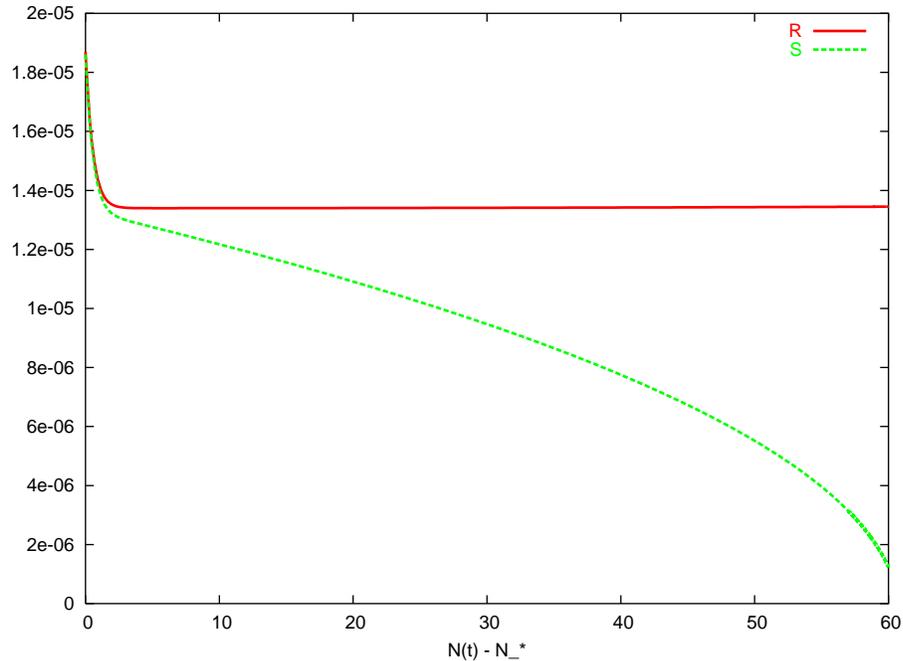}}
\caption{The evolution of the scalar perturbations for a given mode
  $k$ for the case $n=2$. We tune $V_0 = 10^{-10}$ to give the correct
  normalization for 
  ${\mathcal{R}}$. This shows the 60 e--folds after horizon
  exit. Note that we plot $(P_{\mathcal{R}})^{\frac{1}{2}}$ and  $(P_{\mathcal{S}})^{\frac{1}{2}}$. }\label{fig:chaoticobs}
\end{figure}
If one examines the behavior of ${\mathcal{R}}$ and ${\mathcal{S}}$,
we see that it is markedly different from the behavior observed in [\ref{Ashcroft:2002vj}] where we have a run--away potential for $\chi$. 
\begin{figure}[!ht]
\centerline{\scalebox{0.75}{\includegraphics{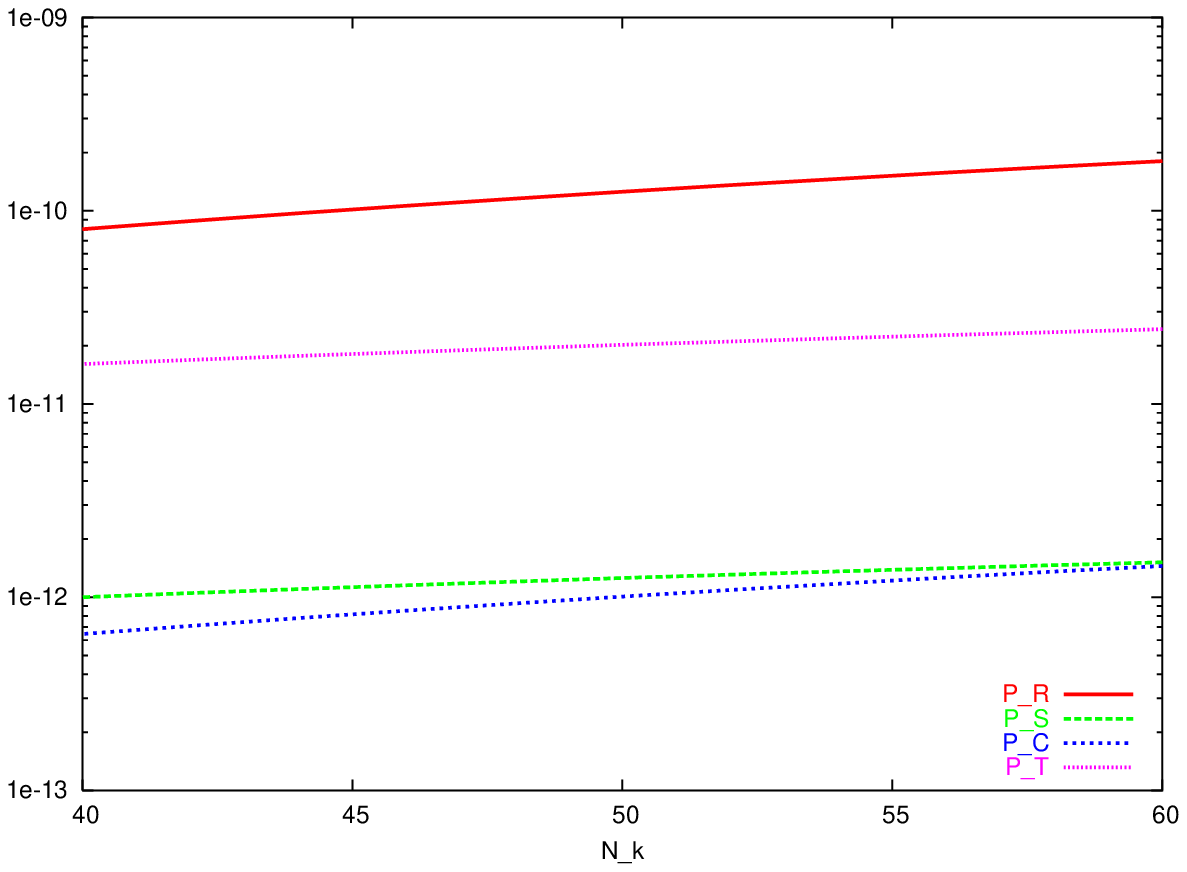}\includegraphics{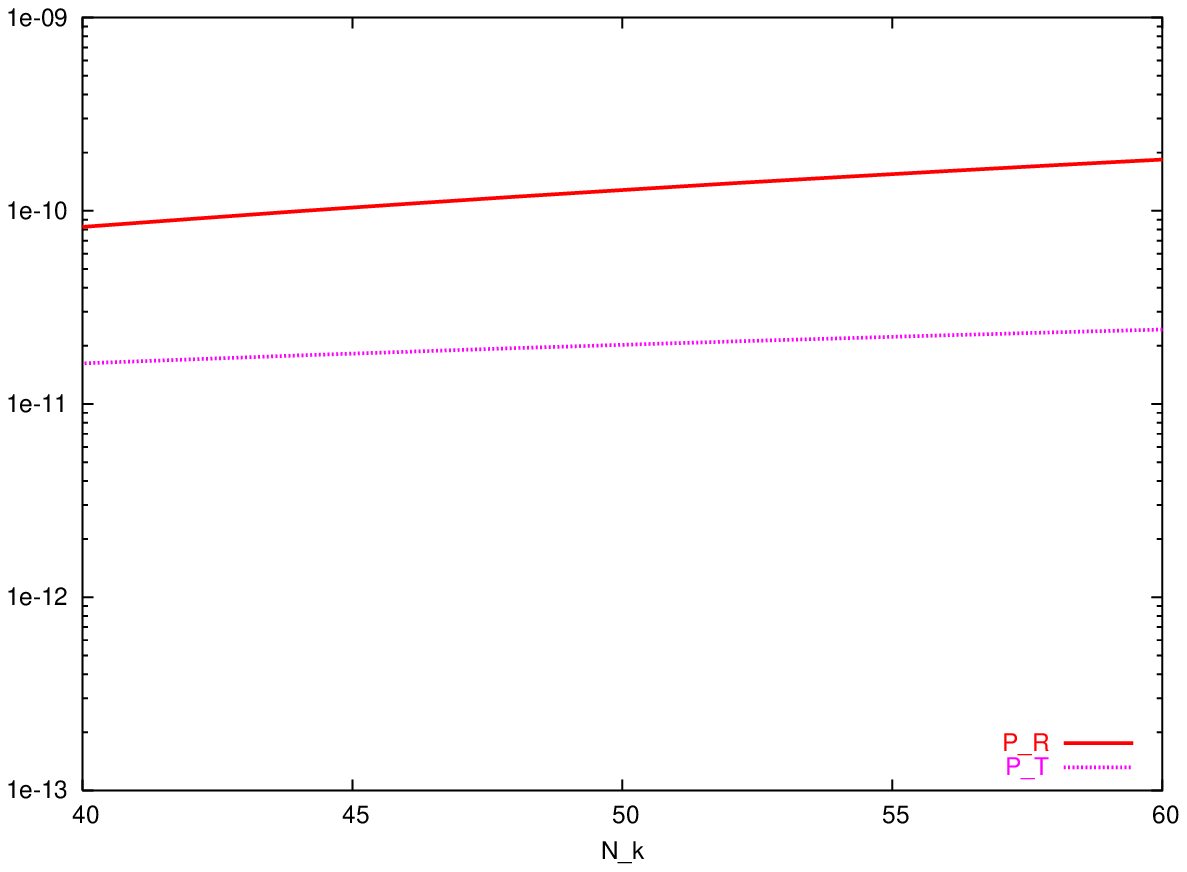}}}
\caption{On the left we plot the spectra for the case $n=2$, $\alpha =
  0.01$ with $m_\chi \sim 10^{-5}$ ( $V_0 = 10^{-10}$ ). We see some running in the
  indices. In contrast, on the right we plot the same case with
  $\alpha = 0$ with the amplitude of $P_{\mathcal{R}}$ scaled to coincide
with the lefthand plot at $N_k = 40$. This gives $m_\chi \sim 10^{-6}$. Note that there is no
entropy production since we are effectively left with one field and,
hence, there is also no correlation in this instance. We see that for this value of $\alpha$
there is very little difference between the plots for the curvature
and tensor spectra.}\label{fig:chaoticspectra}
\end{figure}
Note that the mass required for the inflaton field is $m_\chi \sim
10^{-5}$ which is a few orders of magnitude heavier that one would
normally use in the single field chaotic inflation model. This should
not be surprising as the conformal factor decays during inflation and
so reduces the effective mass of the inflaton.
\begin{figure}[!ht]
\centerline{\scalebox{0.75}{\includegraphics{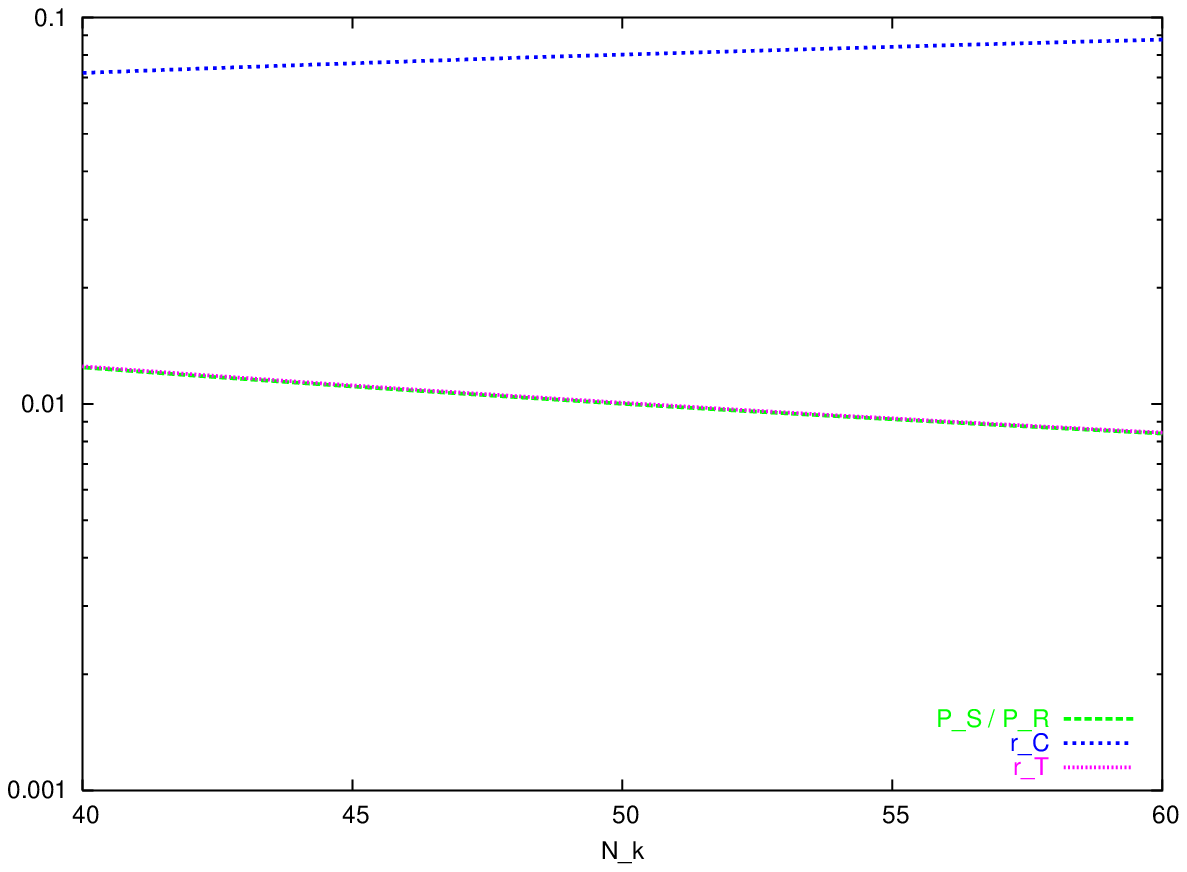}\includegraphics{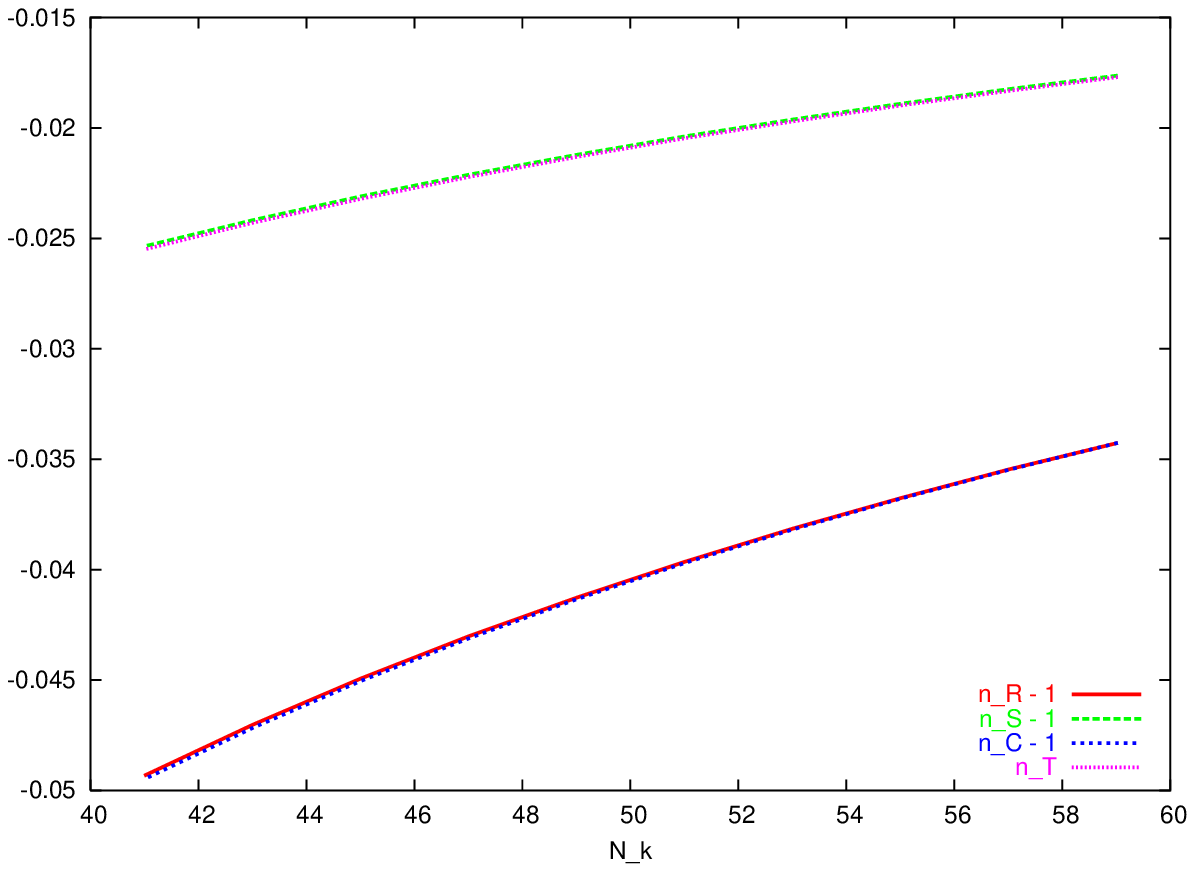}}}
\caption{On the left we plot the evolution of the ratios for $n=2$,
  $\alpha = 0.01$ and $V_0 = 10^{-10}$. Note that whilst the tensor
  production is small, the perturbations are quite well correlated. On the right we show the
  spectral indices. Here we see that the indices for entropy and
  tensor modes are coincident and similarly for the curvature and
  correlation. }\label{fig:chaoticindices}
\end{figure}
Note that the tensor production is small  but that there is a
relatively large correlation between adiabatic and isocurvature
modes. Furthermore we see some running of the indices but their
overall values are consistent with the observations from WMAP. Note
that the indices for entropy and tensor modes are coincident and
similarly with the curvature and correlation. Finally, we find that the ratio 
${\cal P}_{\cal S}/{\cal P}_{\cal R}$ is small, usually of order ${\cal O}(0.01)$.

Although observations demand that 
$\alpha \lesssim 10^{-2}$, it is possible to entertain larger values
provided we stabilize the modulus $\Phi$ with some potential
after the end of inflation. However, the smallness of the slow--roll
parameter $\epsilon_\Phi$ ensures we must have
\begin{eqnarray}
\alpha \ll \frac{1}{\sqrt{2}}.
\end{eqnarray}
to generate slow--roll inflation. One may then wonder what the effect of changing $\alpha$ has on the
indices. For $n=2$, one can see that there is little effect for $\alpha
\leq 10^{-2}$ but as it increases there is significant deviation from
the single field case. In fact the largest possible value which
generates enough inflation is $\alpha = 0.25$ which produces
$n_{\mathcal{R}} - 1 \approx -0.25$.
\begin{figure}[!ht]
\centerline{\scalebox{0.75}{\includegraphics{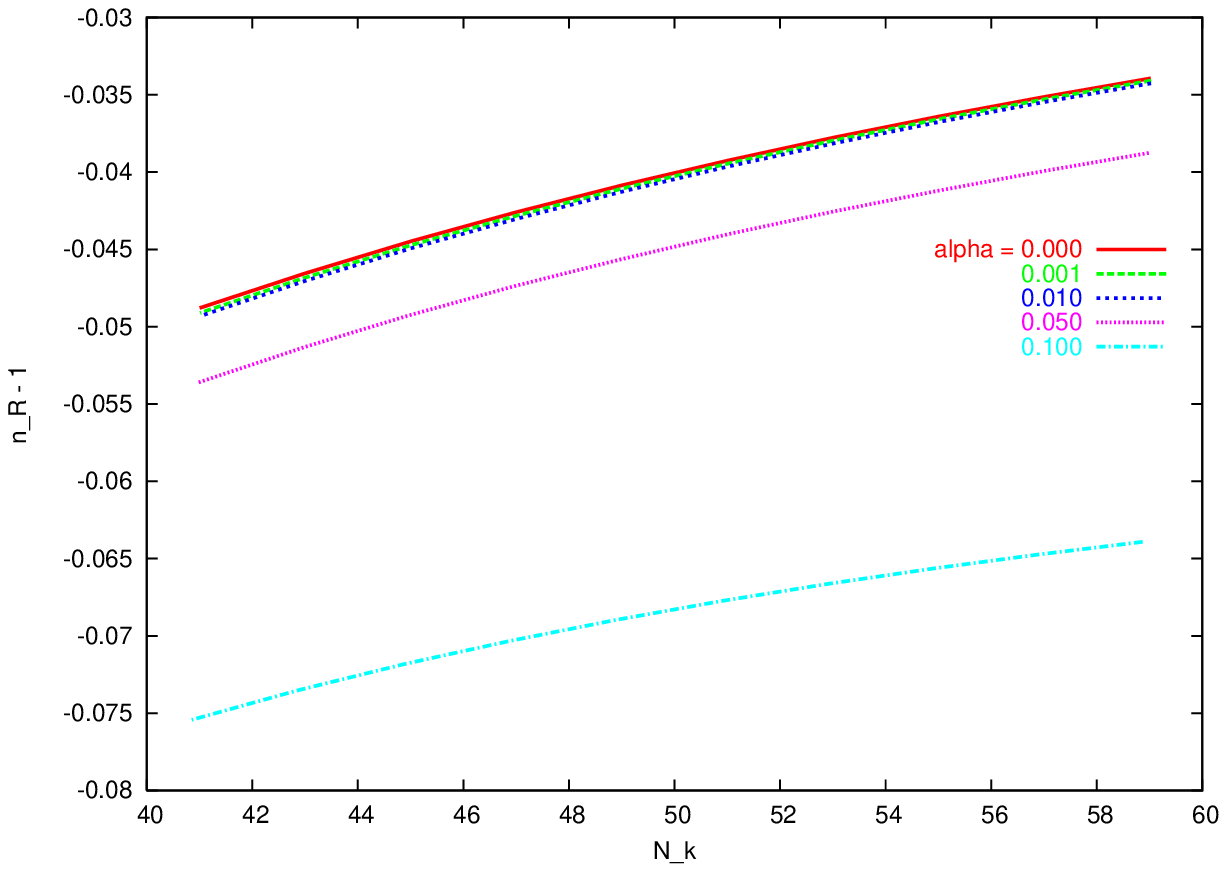}\includegraphics{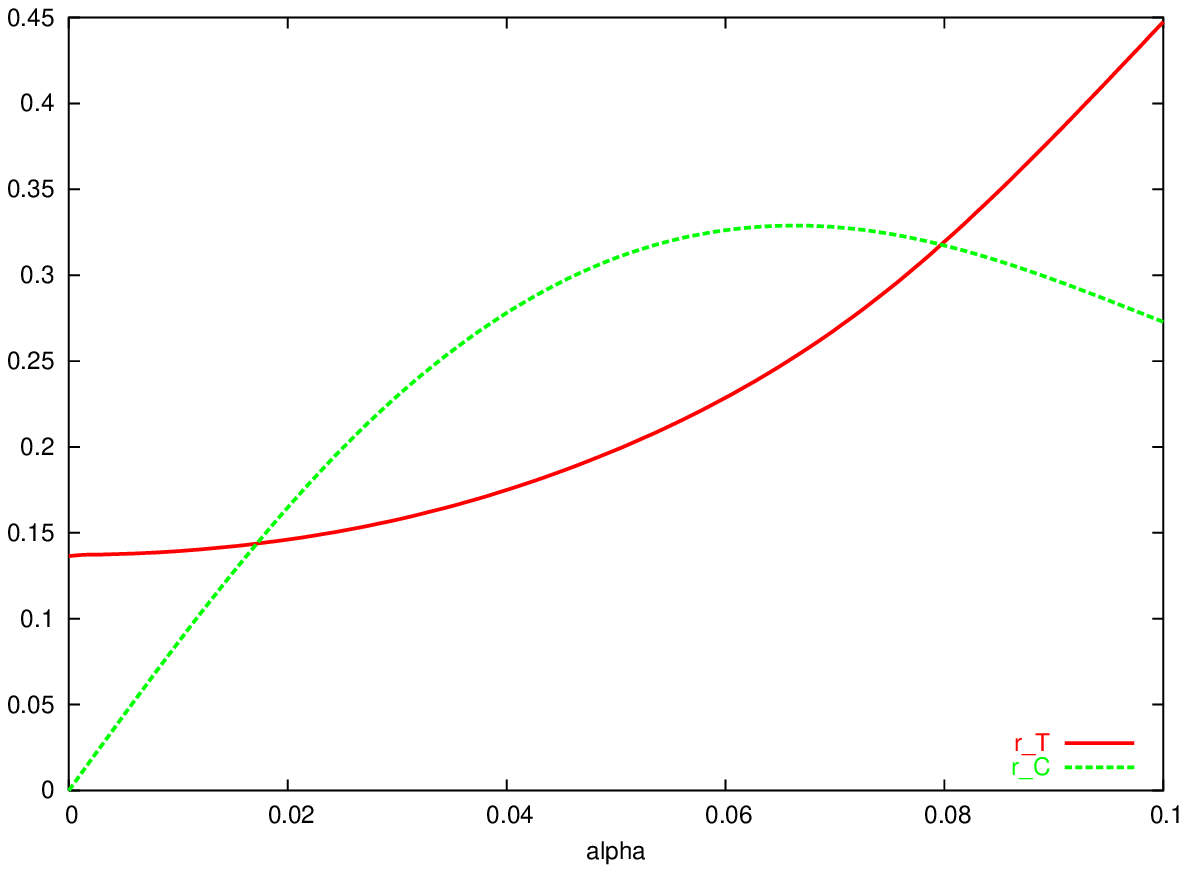}}}
\caption{For the potential $V = \frac{1}{2} m_\chi^2 \chi^2$, as we increase $\alpha$ we are drawn away from
  scale--invariance. Furthermore we see that this also leads to
  increased tensor production, scaled to agree with the normalization
  of WMAP. In addition it is possible to generate strong correlation
  between the perturbations. }\label{fig:phi2var}
\end{figure}
This is not compatible with WMAP,  [\ref{Peiris:2003ff}],
which constrains $\alpha \lesssim 0.2$. The
reader is reminded that in the $\alpha \rightarrow 0$ limit the
modulus $\Sigma$ decouples and we are left with a situation equivalent
to the usual chaotic inflation [\ref{Linde:1983gd}].

\begin{figure}[!ht]
\centerline{\scalebox{0.75}{\includegraphics{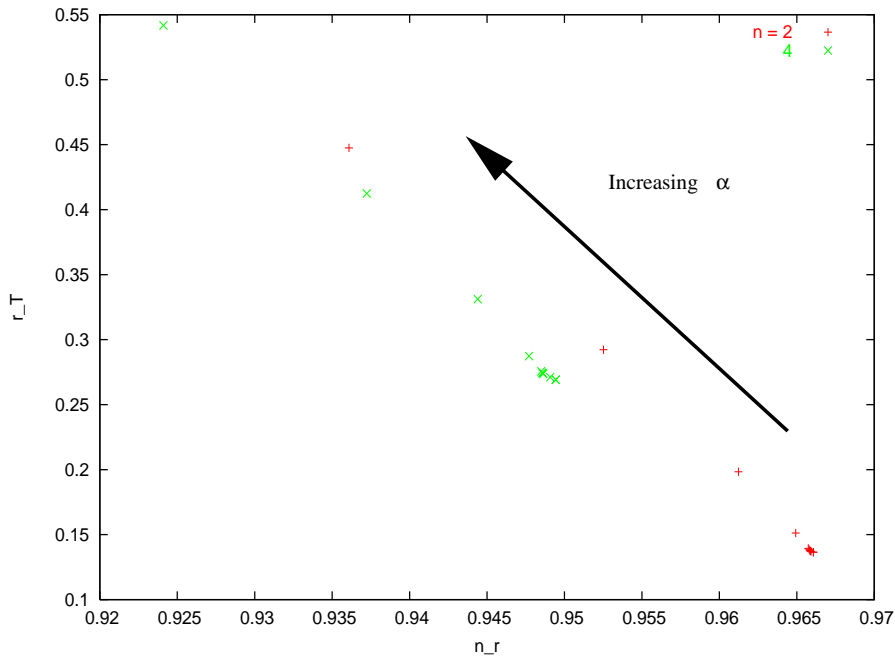}}}
\caption{We see that the increase of $\alpha$
  takes both models away from scale--invariance and leads to greater
  tensor production. This figure was generated with $V_0 = 10^{-10}$
  for both sets of data but is independent of $V_0$.}\label{fig:allowed}
\end{figure}

Recent data from WMAP suggest that the quartic potential is
under strong pressure. This very much depends on the number of e--folds we
take as being observable and on the combination of data sets one 
uses [\ref{Kinney:2003uw}]. Taking $N = 60$, it is still permitted by the
data. As this model is attractive from a particle physics point of view,
we may wish to see if the coupling to the moduli helps to rule in or
out this potential. 
\begin{figure}[!ht]
\centerline{\scalebox{0.75}{\includegraphics{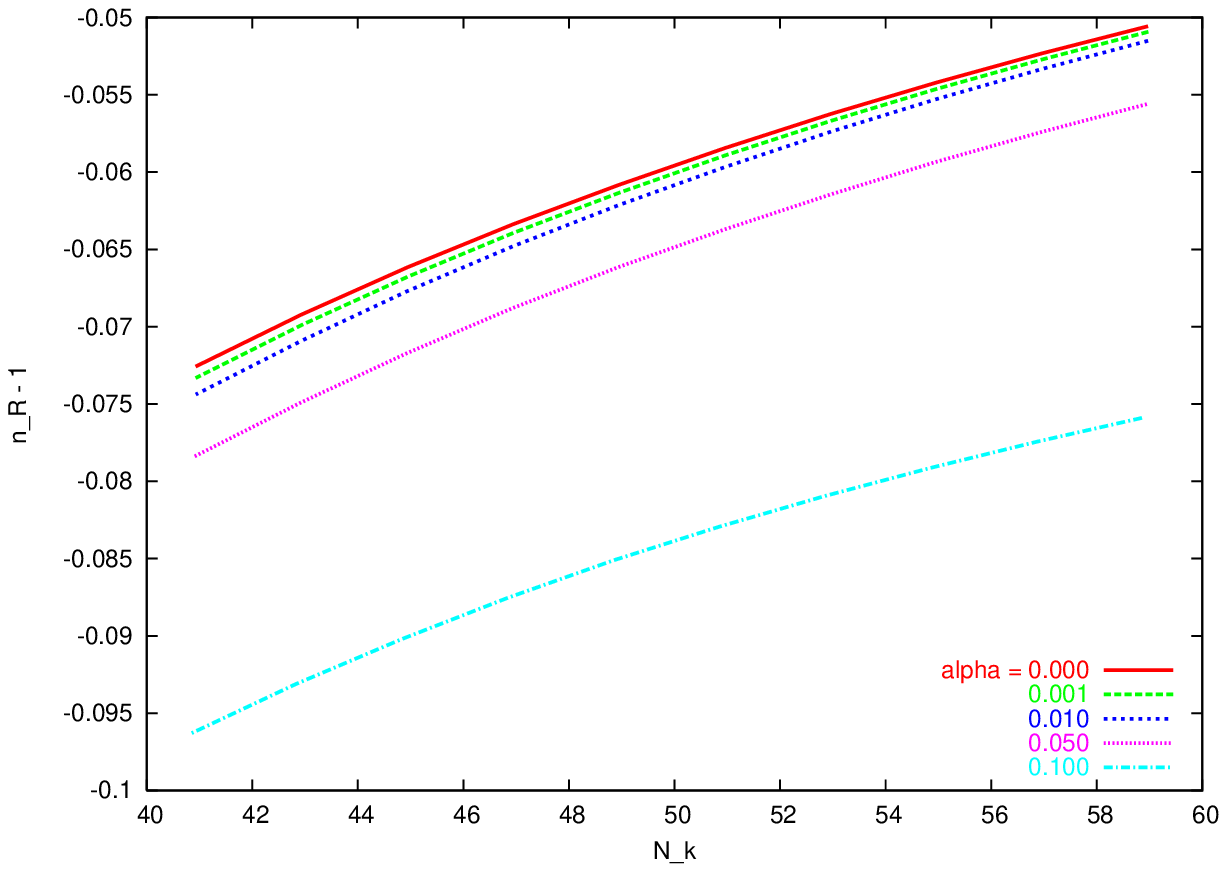}\includegraphics{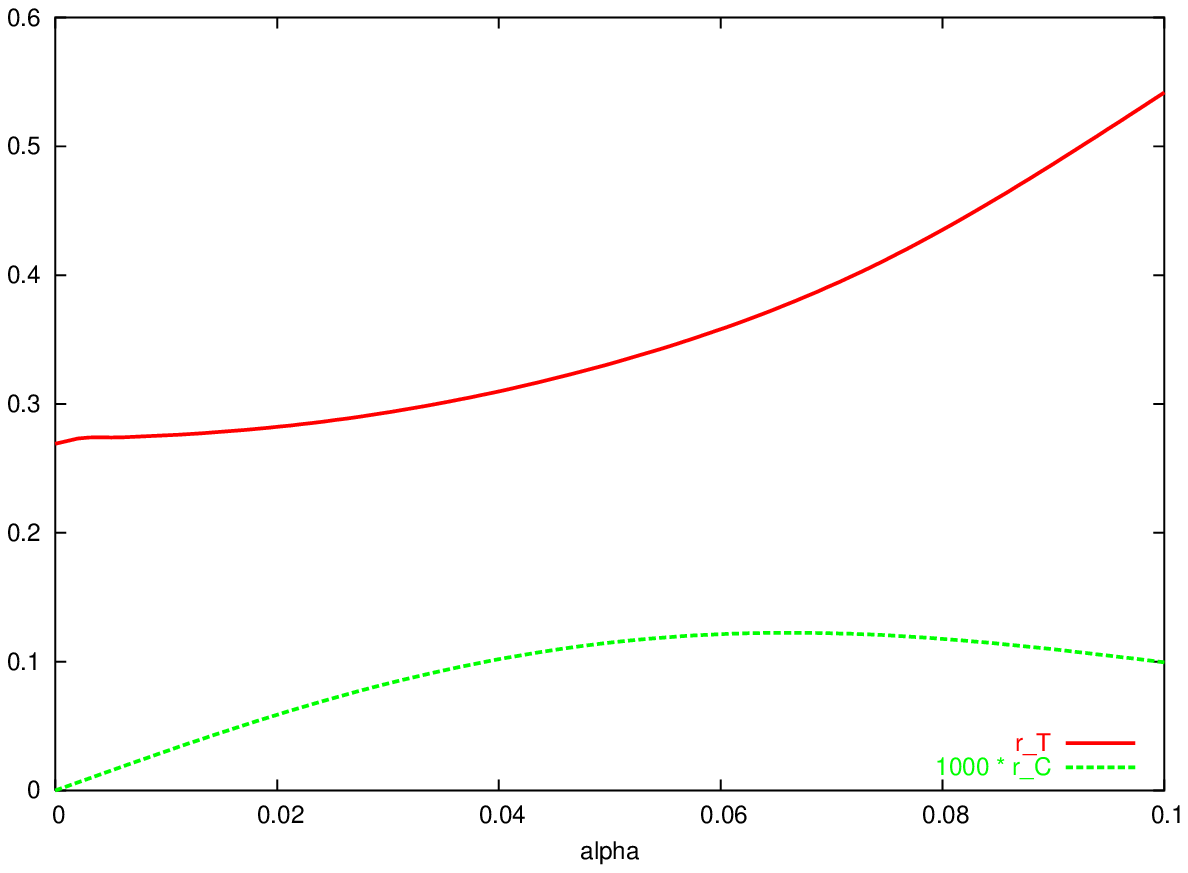}}}
\caption{ In left plot we see the effect of  $\alpha$ on
  $n_{\mathcal{R}}$ for a potential of the form $V_0 \chi^4$. One can
  immediately see that the results are similar for the chaotic
  inflation potential. Moreover, the increase in $\alpha$ also leads
  to an increase in the tensor modes, scaled to agree with the
  normalization of WMAP. The main difference is that these
  perturbations are effectively uncorrelated, $r_{\mathcal{C}} \sim
  10^{-4}$.}\label{fig:effectofalpha}
\end{figure}
One can immediately see from Figure \ref{fig:effectofalpha} that the
coupling to the moduli does us no favors. Increasing $\alpha$
produces more tensor modes-- the main problem associated with
$\chi^4$-- and draws us further away from scale--invariance. A quick
look at Figure \ref{fig:allowed} reveals that we are only just in the
allowed region for $\alpha = 0$. As in the case for the quadratic potential, 
the ratio ${\cal P}_{\cal S}/{\cal P}_{\cal R}$ is small. 

\begin{figure}[!ht]
\centerline{\scalebox{0.75}{\includegraphics{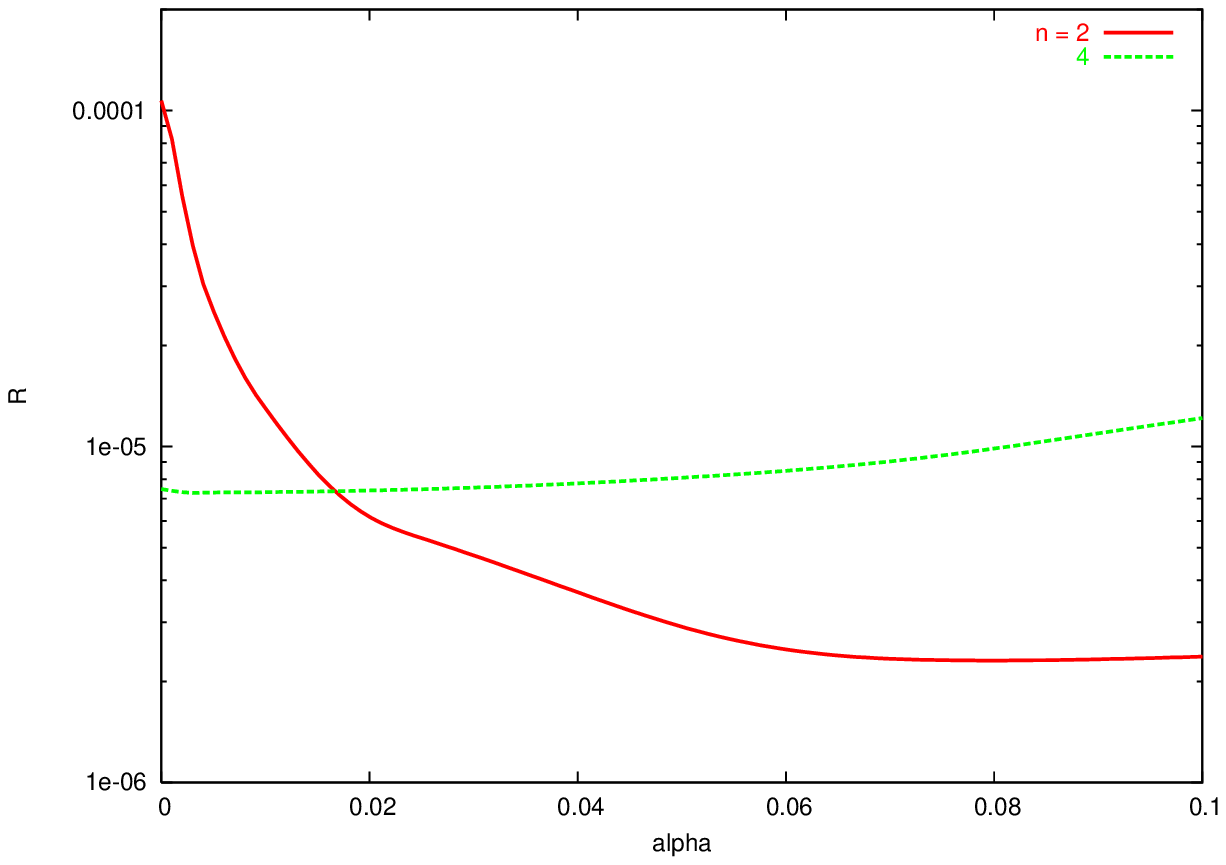}\includegraphics{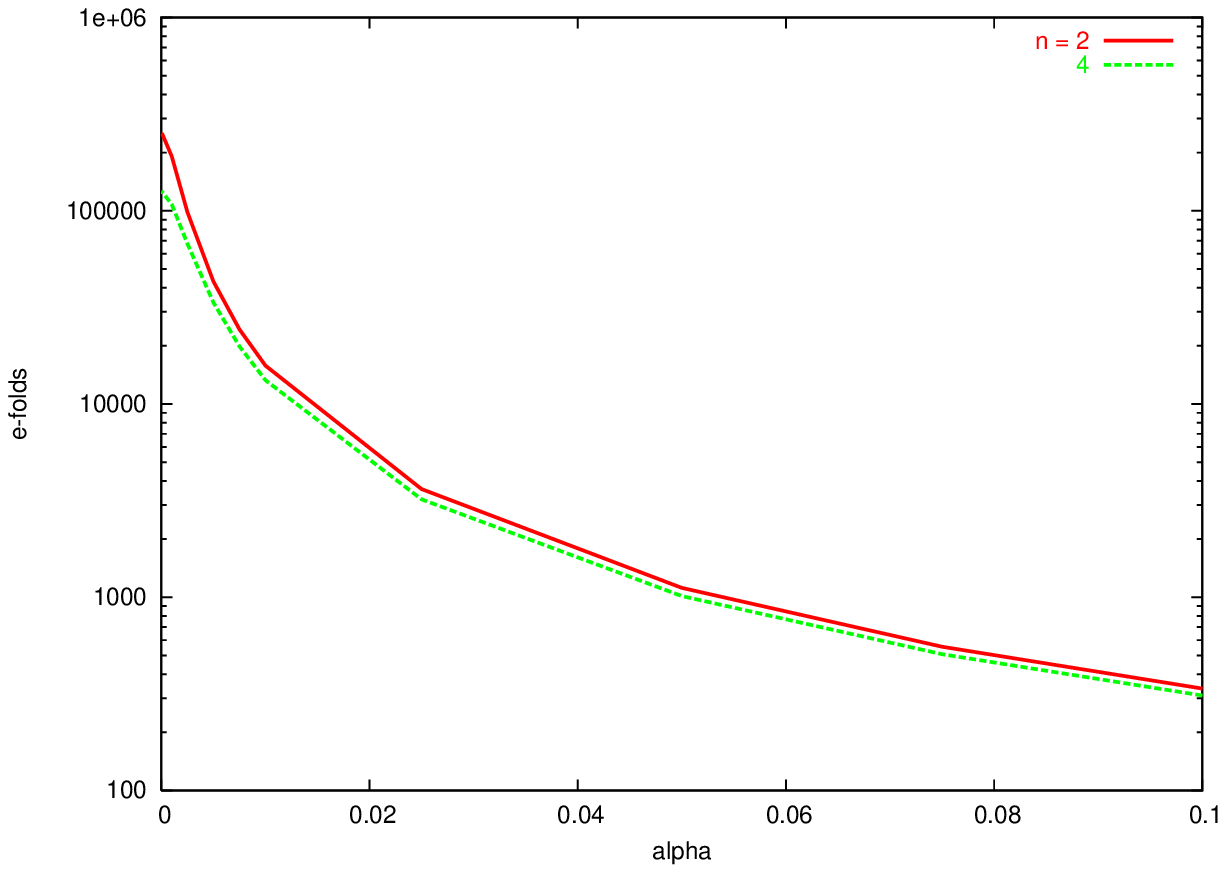}}}
\caption{We examine the effect of $\alpha$ on the amplitude of the
  curvature perturbation. We see that for $n=2$ there is a significant
  effect. We have set $V_0 = 10^{-10}$ at $\alpha = 0.01$ which gives
  $m_\chi \approx 10^{-5}$. Then we see that we get a range of
  $10^{-6} \leq m_\chi \leq 10^{-4}$ for $0 \leq \alpha \leq 0.1$. For
  $n=4$, $\al$ has little effect of the amplitude of $\mathcal{R}$ where
  we set  $V_0 = 10^{-15}$. We also see that reducing $\alpha$ increases the number of e--folds as
  one would expect since the potential becomes less steep.}
\end{figure}

\section{Conclusions}
Boundary inflation in a brane world model with two boundary 
branes and a bulk scalar field was investigated. To study the system, we employed the 
low--energy effective action derived in [\ref{Brax:2002nt}]. Inflation, which must 
have happened at energies much lower than the brane tension if the four--dimensional 
theory is a good description, has some desirable features. 
In particular, the coupling to matter of one of the moduli fields 
(denoted $R$ in this paper) 
is driven towards small values. Therefore, if the inflationary epoch is long enough, the 
coupling of this field is very small after inflation and compatible with observations. 

We found the background solutions for the fields and then studied the
perturbation spectra numerically following the methods of [\ref{DiMarco:2002eb}]. 
We have seen that the effect of the coupling produces spectra which
have indices smaller than one. Furthermore, the amplitude of tensor perturbations is enhanced 
compared to  the predictions based on General Relativity. Our results imply
that the mass of
the inflaton for the quadratic potential can be larger than that normally
required by the single field equivalent. In addition, the adiabatic
and isocurvature perturbations can be highly correlated. In contrast, the 
quartic potential produces highly uncorrelated perturbations with the
coupling having little effect on the required scale of the
potential. Although this model is almost ruled out by current
observations, one may have hoped that the coupling to the moduli may
have helped its cause. We find this not to be the case and, in fact,
the opposite is true. A large coupling would rule out this potential immediately.

Another feature discussed in this paper is that the negative tension brane collapses 
during inflation driven by a scalar field on the positive tension brane. 
Such behavior has already been observed in the matter--dominated era in [\ref{Brax:2002nt}]. 
This suggests that the five--dimensional spacetime is not stable during inflation (and 
matter domination) which is a fundamental problem of brane cosmologies based on models 
with bulk scalars. In the case of the Randall--Sundrum brane world with two branes, 
[\ref{Randall:1999ee}], the negative tension brane is driven towards the AdS horizon, 
whereas the positive tension brane would continue to inflate (and finally enter a 
radiation dominated phase). We would like 
to remind the reader that models based on string or M--theory motivate the existence 
of matter in the bulk (in particular scalar fields). Of course, the results presented 
in this paper will change dramatically once a potential for the fields $\Phi$ and $R$ 
is added. Such a potential could lead to the stabilization of the fields $\Phi$ and $R$ 
thereby avoiding the problem of the 
singularity. However, to derive a potential with cosmologically desirable properties  
is very difficult. In string theory, it is believed that such a potential can 
be derived once the non--perturbative sector of the theory is
understood. Alternatively, string theory may naturally resolve the
problem of the singularity. In our model, the underlying 
field theory has its limitations and the four--dimensional effective theory breaks down
when $R$ approaches zero. In any case, it is clear that work has to be done to understand 
the behavior at $R=0$ better. For example, in [\ref{Brax:2002nt}] it was 
suggested that dark energy is related to this singularity problem. It would also be 
desirable to perform a calculation of the inflationary dynamics and the generation 
of perturbations in the full five--dimensional theory. However, the equations 
governing the perturbations are very difficult to solve. 

Given the fact that models with non--canonical kinetic terms such as the one 
discussed in this paper arise in string theory, our results have an important impact on
inflationary model building. 
At least for the potentials studied here, large couplings (i.e. of ${\cal O}(1)$) 
are not desirable.

\vspace{0.5cm}
\noindent{\bf Acknowledgements:} We are grateful to R.H. Brandenberger
and Ph. Brax for useful discussions. The authors are supported in part 
by PPARC.

%\references
\begin{enumerate}
\item \label{rubakov} V.A. Rubakov, Phys.Usp.{\bf 44}, 871 (2001)
\item \label{branereview} Ph. Brax, C. van de Bruck,  Class.Quant.Grav.{\bf 20}, R201 (2003)
\item \label{langlois} D. Langlois, Prog.Theor.Phys.Suppl.{\bf 148} 181 (2003)
\item \label{wands1} R. Maartens, D. Wands, B.A. Bassett, I. Heard, 
Phys.Rev.D {\bf 62}, 041301 (2000)
\item \label{wands2} D. Langlois, R. Maartens, D. Wands, 
Phys.Lett.B {\bf 489}, 259 (2000)
\item \label{copeland} E.J. Copeland, A.R. Liddle, J.E. Lidsey, 
Phys.Rev.D {\bf 64}, 023509 (2001)
\item\label{Ashcroft:2002ap} P.R. Ashcroft, C. van de Bruck, 
A.C. Davis, Phys. Rev. D. {\bf 66}, 121302 (2002) 
\item \label{lidsey1} G. Huey, J.E. Lidsey, Phys.Rev.D {\bf 66}, 043514 (2002)
\item \label{Liddle} A.R. Liddle, A.J. Smith, Phys.Rev.D {\bf 68}, 
061301 (2003)
\item \label{Liddle2} E. Ramirez, A.R. Liddle, astro-ph/0309608
\item \label{sasaki1} Y. Himemoto, T. Tanaka, M. Sasaki, 
Phys.Rev.D {\bf 65}, 104020 (2002)
\item \label{lidsay2} R.M. Hawkins, J.E. Lidsey, astro-ph/0306311
\item \label{taylor} D. Seery, A. Taylor, astro-ph/0309512
\item \label{linde} A. Linde, Particle Physics and Inflationary Cosmology, 
Harwood Academic Publishers (1990)
\item \label{liddlelyth} A.R. Liddle, D.H. Lyth, Cosmological
  Inflation and Large--Scale Structure, Cambridge University Press (2000) 
\item\label{Brax:2002nt} Ph. Brax, C van de Bruck, A.-C. Davis, C.S. Rhodes, Phys.Rev.D. {\bf 67},023512 (2003)
\item\label{garriga} J. Garriga, O. Pujolas, T. Tanaka, Nucl.Phys. B {\bf 655}, 
127 (2003)
\item\label{Damour:1992we} T. Damour, G. Esposito-Farese, CQG {\bf 9}
  2093 (2001)
\item\label{Rhodes:2003ev} C.S. Rhodes, C. van de Bruck, Ph. Brax,
  A.-C. Davis, astro-ph/0306343
\item\label{transpa} J. Martin, R. H. Brandenberger, Phys.Rev.D{\bf 63},123501 (2001)
\item \label{wands} J. Garcia-Bellido, D. Wands, Phys.Rev.D {\bf 52}, 5636 (1995)
\item \label{wandsbellido} J. Garcia-Bellido, D. Wands, Phys.Rev.D {\bf 52}, 6739 (1995)
\item \label{damour} T. Damour, A. Vilenkin, Phys.Rev.D {\bf 53}, 2981 (1996)
\item\label{Randall:1999ee} L. Randall, R. Sundrum,
  Phys. Rev. Lett.{\bf 83}, 3370 (1999) 
%\item\label{branes1} P. Binetruy, C. Deffayet, D. Langlois,
%  Nucl. Phys. B {\bf 565}, 269 (2000)
%\item\label{branes2} P. Binetruy, C. Deffayet, D. Langlois,
%  Phys. Lett. B {\bf 477}, 285 (2000)
%\item\label{rs2} L. Randall, R. Sundrum, Phys. Rev. Lett.{\bf83},
%  4690 (1999) 
\item\label{brax:2000xk} Ph. Brax, A.-C. Davis Phys. Lett. B {\bf
     497}, 289 (2001)
\item\label{Ashcroft:2002vj} P.R. Ashcroft, C. van de Bruck,
  A.C. Davis, astro-ph/0210597
\item\label{kobayashi}S. Kobayashi, K. Koyama, JHEP {\bf 0212}, 056 (2002)
\item\label{lukas} A. Lukas, B.A. Ovrut, D. Waldram, 
Phys.Rev.D {\bf 61}, 023506 (2000)
\item\label{kofman} A.V. Frolov, L. Kofman, hep-th/0309002
\item\label{DiMarco:2002eb} F. Di Marco, F. Finelli, R. Brandenberger,
  Phys.Rev.D {\bf 67}, 063512 (2003) 
\item\label{Gordon:2000hv} C. Gordon, D. Wands, B.A. Bassett, R. Maartens, 
Phys.Rev.D {\bf 63}, 023506 (2001)
\item\label{Tsujikawa:2002qx} S. Tsujikawa, D. Parkinson,  B.A. Bassett,Phys. Rev.D{\bf 67} 083516 (2003)
\item\label{Peiris:2003ff} H. V. Peiris et al, Astrophys.J.Suppl. {\bf 148}, 213 (2003)
\item\label{Linde:1983gd} A. Linde,Phys. Lett.B {\bf 129} 177 (1983)
\item\label{Kinney:2003uw} W. Kinney, E. Kolb,
  A. Melchiorri,A.Riotto, astro-ph/0305130
\end{enumerate}

\end{document}